\documentclass[5p,times,authoryear]{elsarticle}

\usepackage{ecrc}

\volume{00}

\firstpage{1}

\journalname{Astronomy and Computing}

\runauth{O. Creaner, K. Nolan, E. Hickey \& N. Smith}

\jid{astron.comput.}

\jnltitlelogo{Astronomy and Computing}

\usepackage{amssymb,amsmath}

\usepackage{caption}
\DeclareCaptionType{equ}[Expression][]

\usepackage[figuresright]{rotating}

\usepackage[table]{xcolor}

\usepackage{makecell}

\begin{document}

\begin{frontmatter}



\dochead{}

\title{The Locus Algorithm I: A technique for identifying optimised pointings for differential photometry}


\author[ITTD,DIAS,LBNL]{Ois\'{i}n Creaner\corref{cor1}}
\cortext[cor1]{Corresponding author}
\ead{creanero@gmail.com, oocreaner@lbl.gov}

\author[ITTD]{Kevin Nolan}
\ead{kevin.nolan@tudublin.ie}

\author[ITTD]{Eugene Hickey}
\ead{eugene.hickey@tudublin.ie}

\author[CIT]{Niall Smith}

\address[ITTD]{Technological University Dublin, Tallaght Campus, Dublin 24, Ireland}
\address[DIAS]{Dublin Institute for Advanced Studies, 31 Fitzwilliam Place, Dublin 2, Ireland}
\address[CIT]{Cork Institute of Technology, Bishopstown, Cork, Ireland}
\address[LBNL]{Lawrence Berkeley National Laboratory, 1 Cyclotron Road, Berkeley, California, USA}

\begin{abstract}
Studies of the photometric variability of astronomical sources from ground-based telescopes must overcome atmospheric extinction effects.  Differential photometry by reference to an ensemble of reference stars which closely match the target in terms of magnitude and colour can mitigate these effects.  This Paper describes the design, implementation and operation of a new algorithm, The Locus Algorithm; which enables optimised differential photometry.  The Algorithm is intended to identify, for a given target and observational parameters, the Field of View (FoV) which includes the target and the maximum number of reference stars similar to the target.  A collection of objects from a catalogue (e.g. SDSS) is filtered to identify candidate reference stars and determine a rating for each which quantifies its similarity to the target.  The algorithm works by defining a locus of points around each candidate reference star, upon which the FoV can be centred and include the reference at the edge of the FoV.  The Points of Intersection (PoI) between these loci are identified and a score for each PoI is calculated.  The PoI with the highest score is output as the optimum pointing.  The steps of the algorithm are precisely defined in this paper. The application of The Locus Algorithm to a sample target, SDSS1237680117417115655, from the Sloan Digital Sky Survey is described in detail. The algorithm has been defined here and implemented in software which is available online. The algorithm has also been used to generate catalogues of pointings to optimise Quasar variability studies and to generate catalogues of optimised pointings in the search for Exoplanets via the transit method.  
\end{abstract}


\end{frontmatter}



\section{Introduction}
\label{introduction}

Photometric variability studies involve identifying variations in the brightness of celestial objects as a function of time. Intrinsic variability can occur on timescales from milliseconds to years and in some instances is a critical tool in allowing competing theoretical models to be assessed. For example, time-resolved precision photometry has the potential to infer very small scale structures in astrophysical jets at a scale which is not possible with direct imaging \citep{smith2008emccd}. Alternatively, precision photometry can detect Earth-sized planets around M-type stars via the transit method \citep{giltinan2011using,everett2001technique}.  Variability which is non-repeatable in nature, such as in astrophysical jets, is the most difficult to quantify and places the greatest requirements for the reliability of the photometry being as high as possible. 

The Earth's atmosphere causes incoherent wavelength-dependent variations in the flux detected from a source on timescales from 10ms upwards. If uncorrected, this can appear as intrinsic variability, leading to erroneous conclusions about the structure of a source and the underlying astrophysical drivers \citep{smith2008emccd}. Differential photometry, in which a target is compared to a number of reference stars, attempts to minimise extrinsic effects atmospheric by comparing the brightness of a target to reference stars in the same Field of View (FoV). The assumption is that stars which are spatially close to the target undergo the same atmospheric influences as the target itself \citep{burdanov2014astrokit}. If the stars are of similar brightness and colour as the target, then any effects of the atmosphere distributed across very short spatial scales should equally affect all objects in the FoV \citep{young1991precise,howell2006handbook}. Clearly, the more similar the objects in the FoV used to determine the photometry, the less likely it is that the atmosphere will play a major role in determining the precision. However, photon-limited precisions are rarely achieved in practice, indicating that the removal of atmospheric effects even amongst objects which are spatially close, is limited by the atmospheric conditions at the time of the observations in a complicated way \citep{everett2001technique,howell2002some}.   It has been shown that the general approach to differential photometry can be improved upon by taking very short integrations and selecting those time periods in which the atmosphere is most stable \citep{giltinan2011using}.

Despite the knowledge that reference stars which are similar to the source provide the best approach, there has been no systematic attempt to define the optimum FoV around a target source in which parameters such as colour, magnitude, field-crowding and field orientation have been used to determine the optimum pointings. Here we describe an algorithm, the Locus Algorithm, which identifies the pointing for the which the resultant observational FoV includes the target and the most photometrically appropriate reference stars available.

%

\section{Conceptual basis to The Locus Algorithm}
\label{conceptual-basis-to-the-locus-algorithm}

\begin{figure}[!htb]
\centering
\includegraphics[width=0.47\textwidth]{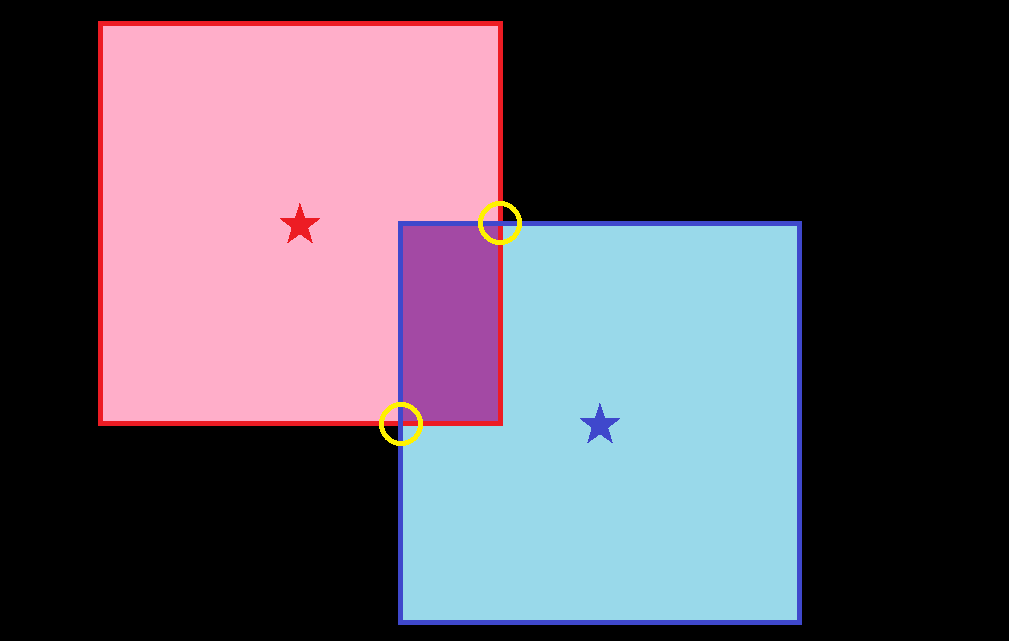}
\caption{\label{loci_concept}Diagrammatic representation of two stars with loci
(red and blue perimeter lines), which intersect and produce two Points
of Intersection (PoI) circled in yellow.  Modified from \citet{creaner2016thesis}}
\end{figure}

A locus can be defined around any star such that a FoV centred on any
point on the locus will include the star at the edge of the FoV. For
fields containing stars close to one another, if one locus intersects
with another, they produce Points of Intersection (PoI) as shown in Figure \ref{loci_concept}.

A FoV centred on any such PoI will include both stars associated with
creating it. At PoI the set of stars that can be
included in a FoV changes.

The Locus Algorithm considers candidate reference stars in what is
termed a Candidate Zone (CZ) - the zone of sky centred on the target
within which a FoV can be selected which includes both the reference
star and the target. For Candidate Reference Stars within the CZ, 
loci are determined, and all relevant PoI are identified. Each PoI 
is assigned a score derived from the number and similarity of reference
stars included in a FoV centred on that PoI. The PoI with the highest score 
becomes the pointing for the target.

\section{Locus Algorithm Design}
\label{locus-algorithm-design}
Based on the conceptual outline above, this section provides a mathematical
definition of the Locus Algorithm and an explanation of the terms used in it.
Section \ref{example-implementation-of-the-locus-algorithm} below describes a
 worked example of this algorithm applied to a sample star, \textit{SDSS1237680117417115655}.

\subsection{Definition of Coordinate System and Locus}
\label{definition-of-coordinate-system-and-locus}

For computational efficiency, The Locus Algorithm considers a Field of
View to be a rectangular area on the sky orientated such that the edges
are aligned with the primary x and y axes of the Cartesian coordinate
system. Movement of the field is restricted to x or y translations.

However, the Celestial coordinate system is defined by the Equatorial
coordinate system, with coordinates specified by Right Ascension (RA)
and Declination (Dec). Because this is a spherical coordinate system,
unit angle in RA is foreshortened, with the degree of foreshortening
defined in Expression \ref{RAshort}
\begin{equ}[!htb]
  \begin{equation}
angle\ in\ RA = {\frac{True\ Angle}{cos(Dec)}}
  \end{equation}
\caption{\label{RAshort}Right Ascension foreshortening with Declination}
\end{equ}

By using this conversion, it is possible to approximate to a high degree
of accuracy a Cartesian coordinate system using RA and Dec; with a small
FoV of East-West size \(R\) and North-South size \(S\) about a target located at
point \(RA_t\) and \(Dec_t\).  Expression \ref{Rprime} defines a corrected angular size in RA direction ($R^\prime$)

\begin{equ}[!htb]
  \begin{equation}
R^\prime = {\frac{R}{cos(Dec_t)}}
  \end{equation}
\caption{\label{Rprime}Definition of a corrected angular size along the RA direction (R$^\prime$)}
\end{equ}

Given these terms, Expression \ref{FoVDef} defines the FoV.

\begin{equ}[!htb]
\begin{equation}
\begin{split}
&RA_t - {\frac{R^\prime}{2}} \leq RA \leq RA_t + {\frac{R^\prime}{2}} \\
&Dec_t - {\frac{S}{2}} \leq Dec \leq Dec_t + {\frac{S}{2}}
\end{split}
\end{equation}
\caption{\label{FoVDef}Definition of a FoV of size R x S centred on a target at
(\(RA_t\) , \(Dec_t\))}
\end{equ}

This definition is accurate to approximately 1\% for a FoV of area 15 $'${} square outside celestial polar regions as shown in Expression \ref{polar}.

\begin{equ}[!htb]
  \begin{equation}
\begin{split}
&Given: R, S = 15' , \lvert Dec\rvert \leq 66.5^{\circ} \\
&{\frac{\lvert\frac{R}{cos(Dec-S)}-\frac{R}{cos(Dec+S)}\rvert}{\frac{R}{cos(Dec)}}}\leq 0.01
\end{split}
  \end{equation}
\caption{\label{polar}Evaluation of the accuracy of the R$^\prime$ for areas away from the celestial pole.}
\end{equ}

We can therefore define the locus about any star on the sky located at
\(RA_t\) and \(Dec_t\) as the values of Right Ascension and Declination
as defined in Expression \ref{FoVDef}.

\subsection{Candidate Zone}
\label{candidate-zone}

A Candidate Zone is defined as a region centred on the target, equal to
four times the area of the Field of View, within which any
reference star can be included in a Field of View with the target and
can therefore be considered as a candidate reference star in identifying
the optimum pointing. Conversely, stars outside the candidate zone
cannot be included in a Field of View with the target and cannot
therefore be considered as candidates reference stars. Hence the
Candidate Zone is the maximum region of sky centred on the target from
which to choose candidate reference stars when identifying an optimum
pointing for a given target. For a target positioned at coordinates
\(RA_t\) and \(Dec_t\) the resulting Candidate Zone is defined by Expression \ref{CZdef}.
\begin{equ}[!htb]
  \begin{equation}
\begin{split}
&RA_t - R^\prime \leq RA_r \leq RA_t + R^\prime \\
&Dec_t - S \leq Dec_r \leq Dec_t + S
\end{split}
  \end{equation}
\caption{\label{CZdef}Definition of a Candidate Zone of size 2R x 2S centred on a
target with coordinates (\(RA_t\), \(Dec_t\)), in which zone reference stars with coordinates (\(RA_r\), \(Dec_r\)) can be found.}
\end{equ}


\subsection{Identification and Filtering of Reference Stars}
\label{identification-and-filtering-of-reference-stars}

For each target, a list of candidate reference stars in its Candidate
Zone is produced based on the following criteria:

\textbf{Position}: the reference star must be in the Candidate Zone as defined in Expression \ref{CZdef}.

\textbf{Magnitude}: the magnitude of the reference star (\(mag_r\)) must be within a user-defined limit (\(\Delta mag\)) of the target's magnitude (\(mag_t\)) as shown in Expression \ref{maglim}.

\textbf{Colour}: the colour index (e.g. \(g-r\)) of the reference star (\(col_r\)) must match the colour of the target (\(col_t\)) to
  within a user-specified limit(\(\Delta col\)) as shown in Expression \ref{collim}.

\textbf{Resolvability}: the reference star must be resolvable, i.e. no other
  star that would impact a brightness measurements within a
  user-specified resolution limit.

All stars in the Candidate Zone which pass these initial filters become
the list of candidate reference stars for which loci will be
identified.

\begin{equ}[!htb]
  \begin{equation}
\begin{split}
&mag_t - \Delta mag < mag_r < mag_t + \Delta mag \\
\end{split}
  \end{equation}
\caption{\label{maglim}Definition of the limits of mag difference between the target and references.}
\end{equ}

\begin{equ}[!htb]
  \begin{equation}
\begin{split}
&col_t - \Delta col < col_r < col_t + \Delta col \\
\end{split}
  \end{equation}
\caption{\label{collim}Definition of the limits of colour difference between the target and references.}
\end{equ}

For each star which passes these filters, a \textit{rating} is calculated.  The calculation is a modular element of the algorithm, and can be modified to suit the needs of a given observer. The rating system gives a measure of how close, spectrally, each reference star is to the target as shown in Expression \ref{rating_def}. these ratings are calculated by first calculating the colour indices for the target and the reference using the next longer-wavelength filter (col\textsubscript{l}) and the next shorter-wavelength filter (col\textsubscript{s}).  The difference between each of these colour indices for the target and the reference is calculated ($\Delta{}col$) and compared with the limit ($\Delta{}col_max$).  The ratio between these values is subtracted from one to get a normalised rating between 0 and 1 (note that as defined in Expression \ref{collim}, $\Delta{}col < \Delta{}col_max$, preventing negative ratings).  The ratings from each of the pairs of colour indices are multiplied together to give the final rating for a Candidate Reference Star.

\begin{equ}[!h]
\begin{align*}
col_{l}&= r-i & col_{s}&= g-r \\
\Delta{}col_{l}&= col_{l,t} -  col_{l,r} & \Delta{}col_{s}&= col_{s,t} -  col_{s,r} \\
Rating_{l}&= 1 - \left | \frac{\Delta{}col_{l}}{\Delta{}col_{max}}\right | & Rating_{s}&= 1 - \left | \frac{\Delta{}col_{s}}{\Delta{}col_{max}}\right | \\
\\
Rating&=Rating_{l}\times{}Rating_{s}  &
\end{align*}
\caption{\label{rating_def}Definition of the scoring system as used in the generation of the Quasar Catalogue. \textit{g}, \textit{r} and \textit{i} are SDSS magnitudes.  \textit{col} refers to colour indices.  Subscript \textit{l} and \textit{s} refer to long- and short-wavelength colour indices respectively.  Subscript \textit{t} refers to the target, while subscript \textit{r} refers to a reference. }
\end{equ}

\subsection{Identifying the Effective Locus for each Candidate Reference Star}
\label{identifying-the-effective-locus-for-each-candidate-reference-star}

\begin{figure}
\center{\includegraphics[width=0.47\textwidth]{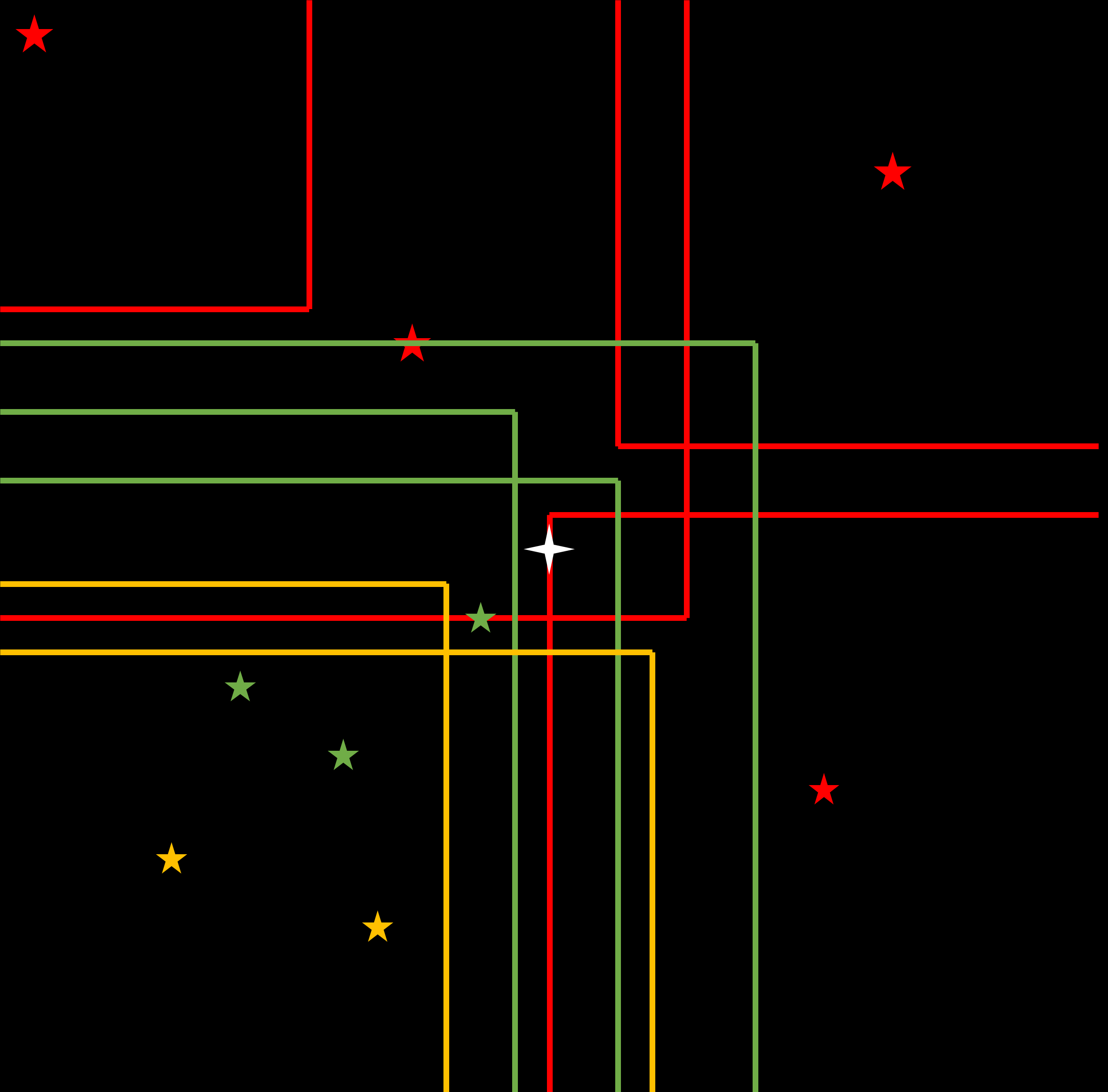}}
\caption{\label{effective_locus}Each effective locus is defined by assigning a pair
of RA and Dec coordinates for a corner point and a pair of lines North
or South and East or West from the corner point. In this diagram, each
candidate reference star is assigned a colour, and the effective locus
that corresponds to it is drawn in the same colour. Modified from \citet{creaner2016thesis}}
\end{figure}

The locus associated with each candidate reference star must be
identified based on Expression \ref{FoVDef}. For the purposes of identifying PoI, only the side surrounding a given candidate reference
star closest to the target need be considered. Hence, we can define the
effective locus for such a candidate reference star as a single line of
constant RA and a single line of constant Dec nearest the target star
as shown in Figure \ref{effective_locus}.

Specifically, the effective locus can be defined as a corner point of
the locus and two lines: one of constant RA and the other of constant
Dec emanating from the corner point. 

Using the Equatorial Coordinate System discussed in Section \ref{definition-of-coordinate-system-and-locus}, with
coordinates of the target specified by \((RA_t, Dec_t)\)
and coordinates of the candidate reference star defined by
\((RA_r, Dec_r)\) and a size of FoV of horizontal
length R and vertical length S, the coordinates of the corner point
\((RA_c, Dec_c)\) are defined as shown in Expression \ref{CPDef}.
\begin{equ}[!htb]
  \begin{equation}
\begin{split}
&RA_t \leq RA_r \Rightarrow RA_c = RA_r- {\frac{R^\prime}{2}}\\
&RA_t > RA_r \Rightarrow RA_c = RA_r+ {\frac{R^\prime}{2}} \\
&Dec_t \leq Dec_r \Rightarrow Dec_c = Dec_r- {\frac{S}{2}}\\
&Dec_t > Dec_r \Rightarrow Dec_c = Dec_r + {\frac{S}{2}}
\end{split}
  \end{equation}
\caption{\label{CPDef}Definition of the corner point (\(RA_c\), \(Dec_c\)) of the effective locus for a FoV of size R x S for a candidate reference star at (\(RA_r\), \(Dec_r\)) and a target at (\(RA_t\), \(Dec_t\)) }
\end{equ}
The directions \(DirRA\) (the direction of the line of constant RA) and 
\(DirDec\) (the direction of the line of constant Dec) of the lines is
determined by the RA and Dec of the candidate
reference star relative to that of the target are given in Expression \ref{DirDef} and as described below.

\begin{itemize}
\item
  If the RA of the candidate is greater than the target, the line of
  constant Dec is defined to be in the direction of increasing RA
\item
  If the RA of the candidate is less than the target, the line of
  constant Dec is defined to be in the direction of decreasing RA
\item
  If the Dec of the candidate is greater than the target, the line of
  constant RA is defined to be in the direction of increasing Dec
\item
  If the Dec of the candidate is less than the target, the line of
  constant RA is defined to be in the direction of decreasing Dec
\end{itemize}

\begin{equ}[!htb]
  \begin{equation}
\begin{split}
&RA_t \leq RA_r \Rightarrow DirDec = +ive\\
&RA_t > RA_r \Rightarrow DirDec = -ive \\
&Dec_t \leq Dec_r \Rightarrow DirRA = +ive\\
&Dec_t > Dec_r \Rightarrow DirRA =-ive
\end{split}
  \end{equation}
\caption{\label{DirDef}Definition the directions (\(DirRA\), \(DirDec\)) of the lines from the corner point of that define the effective locus for a FoV of size R x S for a candidate reference star at (\(RA_c\), \(Dec_c\)) and given a target at (\(RA_t\), \(Dec_t\)).  In current implementations, these values are encoded as a binary switch, with 1 representing increasing (\(+ive\)) direction and 0 representing decreasing (\(-ive\)) direction.}
\end{equ}

\subsection{Identifying and Scoring Points of Intersection}
\label{def_identify}

The points where lines from any two loci intersect are identified and defined as PoI. This involves
comparing the corner point RA and Dec and direction of lines for one
locus with the corner point RA and Dec and direction of lines for a
second locus. In total eight variable associated with each pair of loci are
checked:

\begin{itemize}
\item
  For Locus 1: \(RA_{c1}\), \(Dec_{c1}\), \(DirRA_1\), \(DirDec_1\)
\item
  For Locus 2: \(RA_{c2}\), \(Dec_{c2}\), \(DirRA_2\), \(DirDec_2\)
\end{itemize}

Using these parameters, a check as to whether an intersection between
the two loci occurs is achieved as follows:  

\begin{itemize}
\item
  A line of constant Dec in the positive RA direction from the corner
  point of locus 1 will intersect with a line of constant RA in the
  positive Dec direction from the corner point of locus 2 if locus 1 has
  a lower RA than locus 2 and locus 1 has a higher Dec than locus 2.
\item
  A line of constant RA in the positive Dec direction from the corner
  point of locus 1 will intersect with a line of constant Dec in the
  positive RA direction from the corner point of locus 2 if locus 1 has
  a lower Dec than locus 2 and locus 1 has a higher RA than locus 2.
\end{itemize}

By checking all such possible combinations, all
pairs of loci in the field which result in a PoI are
identified and their RA and Dec noted. The above cases are mathematically 
expressed in Expression \ref{PoICheck}.

\begin{equ}[!htb]
  \begin{equation}
\begin{split}
&if: DirDec_1 = +ive, DirRA_2 = +ive\\
&and: RA_{r1} < RA_{r2}\\
&and: Dec_{r1} > Dec_{r2}\\
&PoI\: exists\: at\: RA_p = RA_{c2}, Dec_p = Dec_{c1}\\
\\
&if: DirRA_1 = +ive, DirDec_2 = +ive\\
&and: RA_{r1} > RA_{r2}\\
&and: Dec_{r1} < Dec_{r2}\\
&PoI\: exists\: at\: RA_p = RA_{c1}, Dec_p = Dec_{c2}\\
\ldots{}
\end{split}
  \end{equation}
\caption{\label{PoICheck}Definition of a PoI (\(RA_p\), \(Dec_p\)) given several sample cases.}
\end{equ}

\begin{figure}[!htb]
\center{\includegraphics[width=0.47\textwidth]{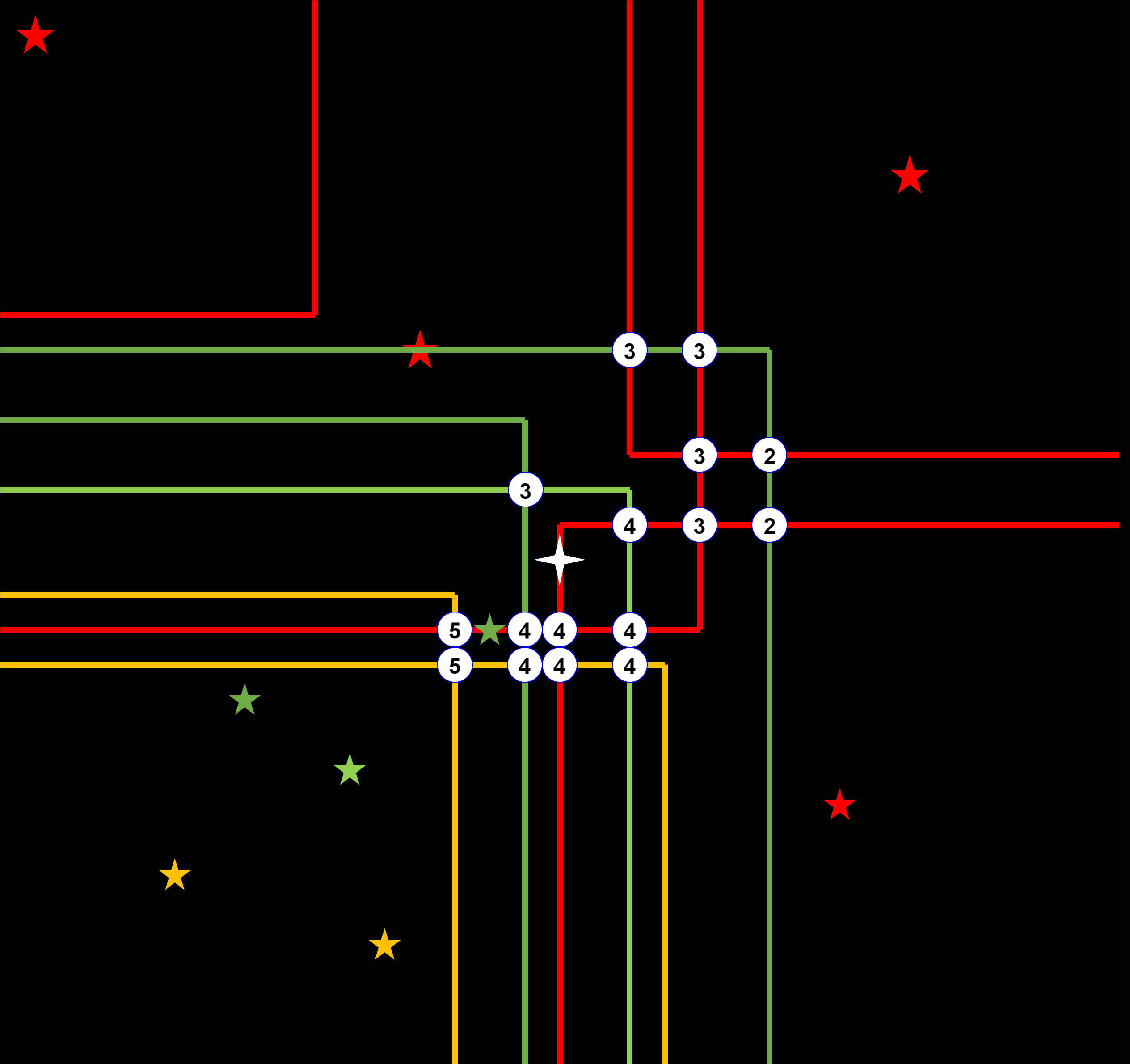}}
\caption{\label{PoIscores}Points of Intersection (PoI), and their associated
score. In this diagram each star has a \textit{rating} of 1, hence the score
associated with each PoI is equal to the number of reference stars
within a FoV centred at that PoI. Modified from \citet{creaner2016thesis}}
\end{figure}

\begin{figure}[!htb]
\center{\includegraphics[width=0.47\textwidth]{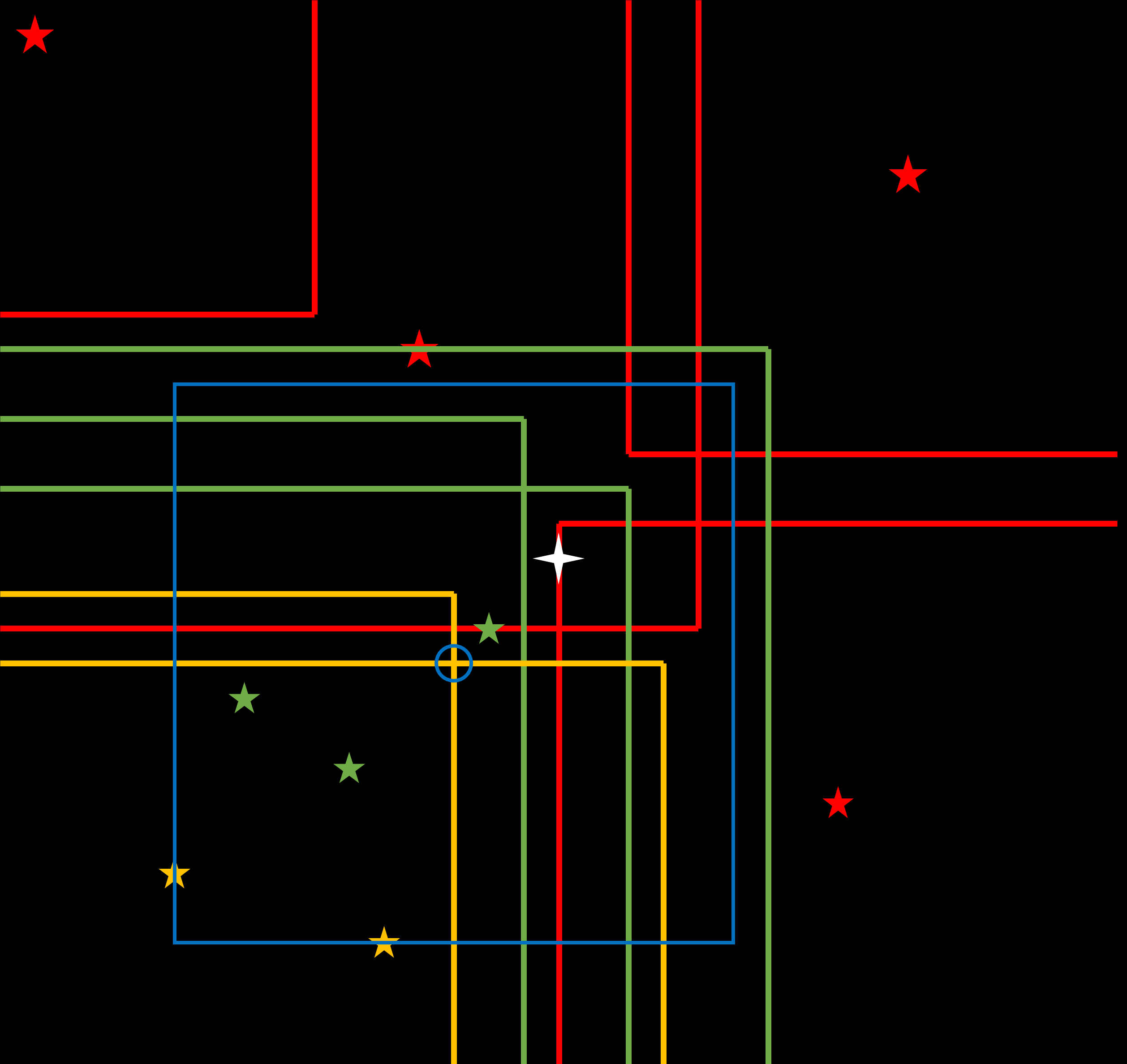}}
\caption{\label{final}Locus Algorithm. Target: white star. Pointing \& FoV:
blue. Reference stars and their loci: Fully in the FoV: greens. On the
edge of the FoV: yellows. Outside FoV: reds.  Modified from \citet{creaner2016thesis}}
\end{figure}

Subsequent to identification, each Point of Intersection is then scored.
This is achieved as follows:

\begin{itemize}
\item
  All of the reference stars which can be included in the FoV centred on the PoI are identified according to Expression \ref{FoVDef}.
\item
  Each reference star has been assigned a \(rating\) between 0 and 1 based
  on its similarity in colour to the target according to Expression \ref{rating_def}.
\item
  The ratings from all counted reference stars in the Field of View are
  added together to give an overall \(score\) for the pointing (See Expression \ref{scoring_def} and Figure \ref{PoIscores}).
\item
  The Point of Intersection with the highest score becomes the pointing for the target (Figure \ref{final}).
\end{itemize}

\begin{equ}[!h]
\begin{align*}
Score&=\sum_{ref}^{FoV}Ratings
\end{align*}
\caption{\label{scoring_def}Definition of the scoring system.  Score is calculated as the sum of all the Ratings for reference stars in the FoV.}
\end{equ}

Scenarios can arise which result in an inability to identify an optimum
pointing for a given target for example if there are no, or a maximum of
one reference stars in the candidate zone; and if no points of
intersection arise -- a scenario which can arise if two (or more)
reference fall in one quadrant of the candidate zone resulting in
concentric loci, or where reference stars are too far apart in different
quadrants of the candidate zone in order for their loci to intersect.
All four of these scenarios are considered in practical implementations
of the Locus Algorithm aimed at identifying the optimum pointings for a
set of targets in a catalogue or list of targets.

In summary, the Locus Algorithm successfully identifies the RA and Dec
coordinates of the optimum pointing for a given target, where optimum
means a field of view with the maximum number of reference stars which
are similar in magnitude and colour to the target.

\section{Example Implementation of the Locus Algorithm}
\label{example-implementation-of-the-locus-algorithm}

To illustrate the operations of the Locus Algorithm, a worked example is
given here. The process described here in producing an optimal pointing
for a given star follows the same sequence of steps described in the
first part of this paper. The process is implemented in the R
programming language and is geared for reproducible research. The code
is available on \citet{githubrepo}
It can be trivially adapted for different target stars and telescope
parameters.

\subsection{Target}
\label{target}

The star \textit{SDSS1237680117417115655}, henceforth called the target,
(${RA = 346.6500^{\circ} ,}\ {DEC = -5.0393^{\circ} }$) is used as the example. This star, in
the constellation Aquarius, has SDSS magnitudes as given on Table \ref{table:target_mags}

\begin{table}[!htb]
\centering
\begin{tabular}{cc}
\hline\hline
Band & SDSS Magnitude\\
\hline
u & 17.20\\
g & 15.38\\
r & 14.65\\
i & 14.40\\
z & 14.28\\
\hline
\end{tabular}
 \caption{SDSS \textit{ugriz} Magnitudes for \textit{SDSS1237680117417115655}, the target used in the sample implementation of the Locus Algorithm}
 \label{table:target_mags} 
\end{table}

The observational parameters are taken from the telescope at
Blackrock Castle Observatory\footnote{CIT Blackrock Castle Observatory, Castle Road, Blackrock, Cork, T12 YW52, Ireland} . This telescope has
parameters given on Table \ref{table:observational_parameters}

\begin{table}[!htb]
\centering
\begin{tabular}{cc}
\hline\hline
Parameters & Values\\
\hline
Field of View in degrees & 0.1667\\
Resolution Limit in degrees & 0.0030\\
Dynamic Range in magnitudes & 2.0000\\
Colour Match Limit & 0.1000\\
\hline
\end{tabular}
 \caption{Observational parameters for the example use of the Locus Algorithm}
 \label{table:observational_parameters} 
\end{table}

\subsection{Candidate Zone}
\label{candidate-zone-1}

The size of the FoV when corrected for shortening by declination is given in expression \ref{Rprime_example}, by
substituting into Expression \ref{Rprime} above.

\begin{equ}[!htb]
  \begin{equation}
  \begin{split}
R^\prime &= {\frac{R}{cos(Dec_c)}} \\
R^\prime &= {\frac{0.1667^{\circ} }{cos(-5.0393^{\circ})}} \\
R^\prime &= 0.16731^{\circ}
\end{split}
  \end{equation}
\caption{\label{Rprime_example}Definition of $R^\prime$ for the target.}
\end{equ}

The locus of the target is given in Expression \ref{locus_example} by substituting into Expression \ref{FoVDef}.

\begin{equ}[!htb]
  \begin{equation}
  \begin{split}
 346.5664^{\circ}  \leq RA \leq 346.7337^{\circ} \\
 -5.1226^{\circ}  \leq Dec \leq -4.9560^{\circ} 
\end{split}
  \end{equation}
\caption{\label{locus_example}Definition of the locus centred on the target.}
\end{equ}

The candidate zone as defined above is the area of sky within which reference stars can possibly be included in the same field of view as the target. This is four times the size of the FoV and is given in Expression \ref{CZ_example} by substituting into Expression \ref{CZdef} above,

\begin{equ}[!htb]
  \begin{equation}
  \begin{split}
346.4827^{\circ}  \leq RA \leq 346.8173 ^{\circ}  \\
-5.2060 ^{\circ}  \leq Dec \leq -4.8726 ^{\circ} 
\end{split}
  \end{equation}
\caption{\label{CZ_example}Definition of the Candidate Zone (CZ) centred on the target.}
\end{equ}

\subsection{Identification and Filtering of Reference Stars}
\label{identification-and-filtering-of-reference-stars-1}

The potential reference stars are selected as follows:

\begin{itemize}
\item
  Position: Within the Candidate Zone defined in Expression \ref{CZ_example}, SDSS records 1345 separate
  objects with clean photometry \citep{aguado2019fifteenth}. These are
  downloaded by an SQL query run on the CAS database, release DR15
\item
  Magnitude: as shown in Expression \ref{maglim} the reference star must be within the dynamic range, 2, of
  the target's magnitude of 14.648, i.e. ${12.648 \leq r \leq 16.648}$. This leaves 41 potential references.
\item
  Colour: as shown in Expression \ref{collim}  the reference star must match the colour of the target to
  within a user-specified limit of 0.1 magnitudes. In this case this means ${0.634 \leq (g-r) \leq 0.834}$ and ${0.149 \leq (r-i) \leq 0.349}$.  This leaves 15 potential references.
\item
  Resolvability: the reference star must be resolvable, i.e. no other
  star that would impact a brightness measurements within a
  user-specified resolution limit, in this case 11$"$ (0.003$^{\circ}$ ). Any object this close to a potential reference star and with
  an \textit{r}-band magnitude which is 5 magnitudes greater than the potential
  reference or brighter will pollute the light from the potential
  reference star. This leaves 14 potential references.
\end{itemize}

These numbers are presented in Table \ref{summary_of_references}, accessed using \citet{rcurl}. 

\begin{table}[!htb]
\centering
\begin{tabular}{cc}
\hline\hline
filters & numbers\\
\hline
Position, in Field of View & 1345\\
Correct Magnitude & 41\\
Correct Colour & 15\\
Resolvable & 14\\
In Final Field of View & 7\\
\hline
\end{tabular}
\caption{\label{summary_of_references}A summary of the number of candidate reference stars remaining at each stage.}
\end{table}

Table \ref{candidate_references} gives the 15 stars in the candidate zone.  14 of these are candidate reference stars.  7 of those 14 are used in the calculation of score for the pointing that is ultimately selected.  Also shown in this table is the target itself, highlighted in green. Table \ref{candidate_references} also includes ratings for each star calculated as per Expression \ref{rating_def}.

\begin{table*}[!htb]
\centering
\begin{tabular}{cccccccccc}
\hline \hline
ref. & objID & ra & dec & u & g & r & i & z & ratings\\
\hline
\rowcolor[HTML]{D7261E}  \textcolor{white}{\textbf{1}} & \textcolor{white}{\textbf{1237680117417115692}} & \textcolor{white}{\textbf{346.7237}} & \textcolor{white}{\textbf{-5.0473}} & \textcolor{white}{\textbf{18.36}} & \textcolor{white}{\textbf{16.58}} & \textcolor{white}{\textbf{15.84}} & \textcolor{white}{\textbf{15.57}} & \textcolor{white}{\textbf{15.46}} & \textcolor{white}{\textbf{0.7345}}\\
2 & 1237680065885241391 & 346.4833 & -4.9462 & 18.97 & 17.14 & 16.42 & 16.16 & 15.99 & 0.8080\\
\rowcolor[HTML]{D7261E}  \textcolor{white}{\textbf{3}} & \textcolor{white}{\textbf{1237680117417115683}} & \textcolor{white}{\textbf{346.7128}} & \textcolor{white}{\textbf{-5.0499}} & \textcolor{white}{\textbf{17.59}} & \textcolor{white}{\textbf{15.78}} & \textcolor{white}{\textbf{15.11}} & \textcolor{white}{\textbf{14.87}} & \textcolor{white}{\textbf{14.80}} & \textcolor{white}{\textbf{0.3610}}\\
4 & 1237680117417050117 & 346.5378 & -5.0185 & 18.40 & 16.52 & 15.77 & 15.53 & 15.39 & 0.7006\\
5 & 1237680117417181202 & 346.8104 & -4.9749 & 17.46 & 15.62 & 14.85 & 14.58 & 14.44 & 0.5353\\
\rowcolor[HTML]{D7261E}  \textcolor{white}{\textbf{6}} & \textcolor{white}{\textbf{1237680065348435996}} & \textcolor{white}{\textbf{346.7072}} & \textcolor{white}{\textbf{-5.1987}} & \textcolor{white}{\textbf{16.70}} & \textcolor{white}{\textbf{14.70}} & \textcolor{white}{\textbf{13.90}} & \textcolor{white}{\textbf{13.68}} & \textcolor{white}{\textbf{13.52}} & \textcolor{white}{\textbf{0.3220}}\\
7 & 1237680065348501526 & 346.8037 & -5.2019 & 15.88 & 14.15 & 13.39 & 13.15 & 12.99 & 0.6440\\
\rowcolor[HTML]{D7261E}  \textcolor{white}{\textbf{8}} & \textcolor{white}{\textbf{1237680117417050120}} & \textcolor{white}{\textbf{346.5626}} & \textcolor{white}{\textbf{-5.1530}} & \textcolor{white}{\textbf{18.46}} & \textcolor{white}{\textbf{16.50}} & \textcolor{white}{\textbf{15.77}} & \textcolor{white}{\textbf{15.53}} & \textcolor{white}{\textbf{15.40}} & \textcolor{white}{\textbf{0.8298}}\\
\rowcolor[HTML]{0D4A05}  \textcolor{white}{\textbf{10}} & \textcolor{white}{\textbf{1237680117417115655}} & \textcolor{white}{\textbf{346.6500}} & \textcolor{white}{\textbf{-5.0393}} & \textcolor{white}{\textbf{17.20}} & \textcolor{white}{\textbf{15.38}} & \textcolor{white}{\textbf{14.65}} & \textcolor{white}{\textbf{14.40}} & \textcolor{white}{\textbf{14.28}} & \textcolor{white}{\textbf{1.0000}}\\
\rowcolor[HTML]{D7261E}  \textcolor{white}{\textbf{11}} & \textcolor{white}{\textbf{1237680117417115762}} & \textcolor{white}{\textbf{346.6755}} & \textcolor{white}{\textbf{-5.1195}} & \textcolor{white}{\textbf{18.92}} & \textcolor{white}{\textbf{17.02}} & \textcolor{white}{\textbf{16.28}} & \textcolor{white}{\textbf{15.97}} & \textcolor{white}{\textbf{15.85}} & \textcolor{white}{\textbf{0.3804}}\\
\rowcolor[HTML]{D7261E}  \textcolor{white}{\textbf{12}} & \textcolor{white}{\textbf{1237680117417050133}} & \textcolor{white}{\textbf{346.5944}} & \textcolor{white}{\textbf{-5.1611}} & \textcolor{white}{\textbf{16.70}} & \textcolor{white}{\textbf{14.83}} & \textcolor{white}{\textbf{14.07}} & \textcolor{white}{\textbf{13.89}} & \textcolor{white}{\textbf{13.65}} & \textcolor{white}{\textbf{0.2410}}\\
13 & 1237680065885241371 & 346.5598 & -4.9312 & 17.68 & 15.93 & 15.23 & 15.01 & 14.90 & 0.4846\\
14 & 1237680065885306903 & 346.6695 & -4.9149 & 17.72 & 15.65 & 14.90 & 14.69 & 14.45 & 0.4292\\
15 & 1237680117417115701 & 346.7433 & -5.0054 & 18.46 & 16.41 & 15.58 & 15.32 & 15.16 & 0.0683\\
\hline
\end{tabular}
\caption{\label{candidate_references}Details of the candidate reference stars from SDSS.  The
seven stars highlighted in red are found in the final field of view. The target star itself
is shown highlighted in green.  Then numbers given in the first column ``ref'' are used to identify the candidate reference stars in the worked examples shown in Subsection \ref{example_loci}.  Data  accessed using RCurl by \citet{rcurl}, table generated with knitr by \citet{knitr1,knitr2,knitr3} and kableExtra by \citet{kableExtra}}
\end{table*}

For example, the star \textit{SDSS1237680117417050120} (Star \#{}8 on Table \ref{candidate_references}) has
${g_r = 16.498}$, ${r_r = 15.771}$, and $i_r = 15.533$. This compares
to the target magnitudes of ${g_t = 15.382}$, ${r_t = 14.648}$, and
${i_t = 14.399}$.  Substituting these into Expression \ref{rating_def} gives the calculation demonstrated in Expression \ref{rating_example}.


\begin{equ}[!h]
\begin{align*}
&col_{l,r}&= r_r-i_r= 15.771-15.533&= 0.238 &\\
&col_{l,t}&= r_t-i_t= 14.648-14.399&= 0.249 &\\
&\Delta{}col_{l}&= col_{l,r} -col_{l,t}= 0.249 -  0.238&= 0.011  &\\
&Rating_{l}&= 1 - \left | \frac{\Delta{}col_{l}}{\Delta{}col_{max}}\right |= 1 - \left | \frac{0.011}{0.1}\right |&= 0.89 &\\
\\
&col_{s,r}&= g_r-r_r= 16.498-15.771&= 0.727 &\\
&col_{s,t}&= r_r-i_r= 115.382-14.648&= 0.734 &\\
&\Delta{}col_{s}&= col_{s,r} -col_{s,t}= 0.727 -  0.734&= -0.007 &\\
&Rating_{s}&= 1 - \left | \frac{\Delta{}col_{s}}{\Delta{}col_{max}}\right |= 1 - \left | \frac{-0.007}{0.1}\right |&= 0.93 &\\
 \\
&Rating&= Rating_{l}\times{}Rating_{s} \\
&&= 0.89\times{}0.93&= 0.83 &
\end{align*}
\caption{\label{rating_example}Definition of the rating of Star \#{}8}
\end{equ}

\subsection{Identifying the Effective Locus for each Candidate Reference Star}
\label{example_loci}

Expression \ref{CPDef} defines how the corner points for a given reference can be calculated.  Substituting in for   Star \#{}8, the resulting corner point is calculated as shown in Expression \ref{CP_example}

\begin{equ}[!htb]
  \begin{equation}
  \begin{split}
	if:& RA_r < RA_t \\
	&\implies RA_c = RA_r + \frac{R^\prime}{2}  \\
	because:& 346.5626^{\circ}  < 346.6500^{\circ}  \\
	&\implies RA_{c,8} = 346.5626^{\circ}   + 0.0837^{\circ} \\ 
	&RA_{c,8} =346.6463^{\circ}   \\
	and\\
	if:& Dec_r < Dec_t \\
	&\implies Dec_c = Dec_r + \frac{S}{2} \\
	because:& -5.1530^{\circ}  < -5.0393^{\circ}  \\
	&\implies Dec\textsubscript{c,8} = -5.1530^{\circ}  + 0.0833^{\circ}  \\
	&Dec\textsubscript{c,8} = -5.0697^{\circ} 
  \end{split}
    \end{equation}
\caption{\label{CP_example}Definition of the corner-point of the effective locus for Reference Star 8}
\end{equ}

From Expression \ref{DirDef}, the direction the locus must be drawn from the corner point to generate the locus can be determined.  Applying this to  Star \#{}8 gives the values shown in Expression \ref{Dir_example}

\begin{equ}[!htb]
  \begin{equation}
  \begin{split}
&if: RA_{r} < RA_{t} \implies DirDec = -ive\\
&because: 346.5626^{\circ}  < 346.6500^{\circ}  \implies DirDec_{8} = -ive\\
&and\\ 
&if: Dec_{r} < Dec_{t} \implies DirRA = -ive\\
&because: -5.1530^{\circ}  < -5.0393^{\circ}  \implies DirRA_{8} = -ive
  \end{split}
    \end{equation}
\caption{\label{Dir_example}Definition of the directions of the lines drawn from the corner-point of the effective locus for Star \#{}8}
\end{equ}

Repeating this calculation for \textit{SDSS1237680065348435996} (Star \#{}6 from Table \ref{candidate_references}) gives the values shown on Table \ref{example_targets}.  These two reference stars are used to calculate Points of Intersection (PoI) in Subsection \ref{PoI_example}

\begin{table}[!htb]
\centering
\begin{tabular}{ccc}
\hline\hline
\textbf{quantity} & \textbf{Ref \#{}6} & \textbf{Ref \#{}8}\\
\hline 
$RA_c$ & $346.6235^{\circ} $ & $346.6463^{\circ} $ \\
$Dec_c$ & $-5.1153^{\circ} $ & $-5.0697^{\circ} $ \\
$DirRA$ & $-ive$ & $-ive$ \\
$DirDec$ & $+ive$ & $-ive$ \\
\hline
\end{tabular}
\caption{\label{example_targets}A summary of the defining parameters of the effective loci for example  Stars \#{}6 and \#{}8}
\end{table}

\subsection{Identifying and Scoring Points of Intersection}
\label{PoI_example}

We can compare these two effective loci to see if they intersect to form a valid PoI. 
Expression \ref{PoICheck} shows some examples of how a valid PoI can be determined.  Applying this process to Stars \#{}6 and \#{}8 gives Expression \ref{PoI_example_expr}:

\begin{equ}[!htb]
  \begin{equation}
  \begin{split}
  &DirRA_{8} =+ive,\ DirDec_{6}=-ive \\
  &RA_{r,8} < RA_{r,6}\ and\ Dec_{r,8} > Dec_{r,6} \\
  &\implies \\
&PoI\ exist\ at :\\
&(RA_p, Dec_p) = (RA_{c,6}, Dec_{c,8})\\
&(RA_p, Dec_p) = (346.6463^{\circ} , -5.1153^{\circ} )\\
  \end{split}
    \end{equation}
\caption{\label{PoI_example_expr}Definition of the PoI between the effective Loci of Stars \#{}6 and \#{}8}
\end{equ}

From Expression \ref{FoVDef}, the FoV centred on the PoI ${(346.6463^{\circ} ,\ -5.1153^{\circ} )}$ can be calculated to be as shown in Expression \ref{FoV_example}
\begin{equ}[!htb]
  \begin{equation}
  \begin{split}
  &RA_p - {\frac{R^\prime}{2}} \leq RA \leq RA_p + {\frac{R^\prime}{2}} 
  \\  & \implies 346.6463^{\circ}  - 0.0837^{\circ}  \leq RA \leq 346.6463^{\circ}  + 0.0837^{\circ}  \\
 &346.5626^{\circ}  \leq RA \leq 346.7299^{\circ}  \\
 &and: \\
  &Dec_p - {\frac{S}{2}} \leq Dec \leq Dec_p + {\frac{S}{2}} \\
  &-5.1153^{\circ}  - 0.0833^{\circ}  \leq Dec \leq -5.1153^{\circ}  + 0.0833^{\circ}  \\
  &-5.1987^{\circ}  \leq Dec \leq -5.0320^{\circ} 
  \end{split}
    \end{equation}
\caption{\label{FoV_example}Definition of the FoV centred on the PoI between the effective Loci of Stars \#{}6 and \#{}8}
\end{equ}  

\begin{figure}[!htb]
\center{\includegraphics[trim={0 9.75cm 0 6cm},clip,width=0.47\textwidth]{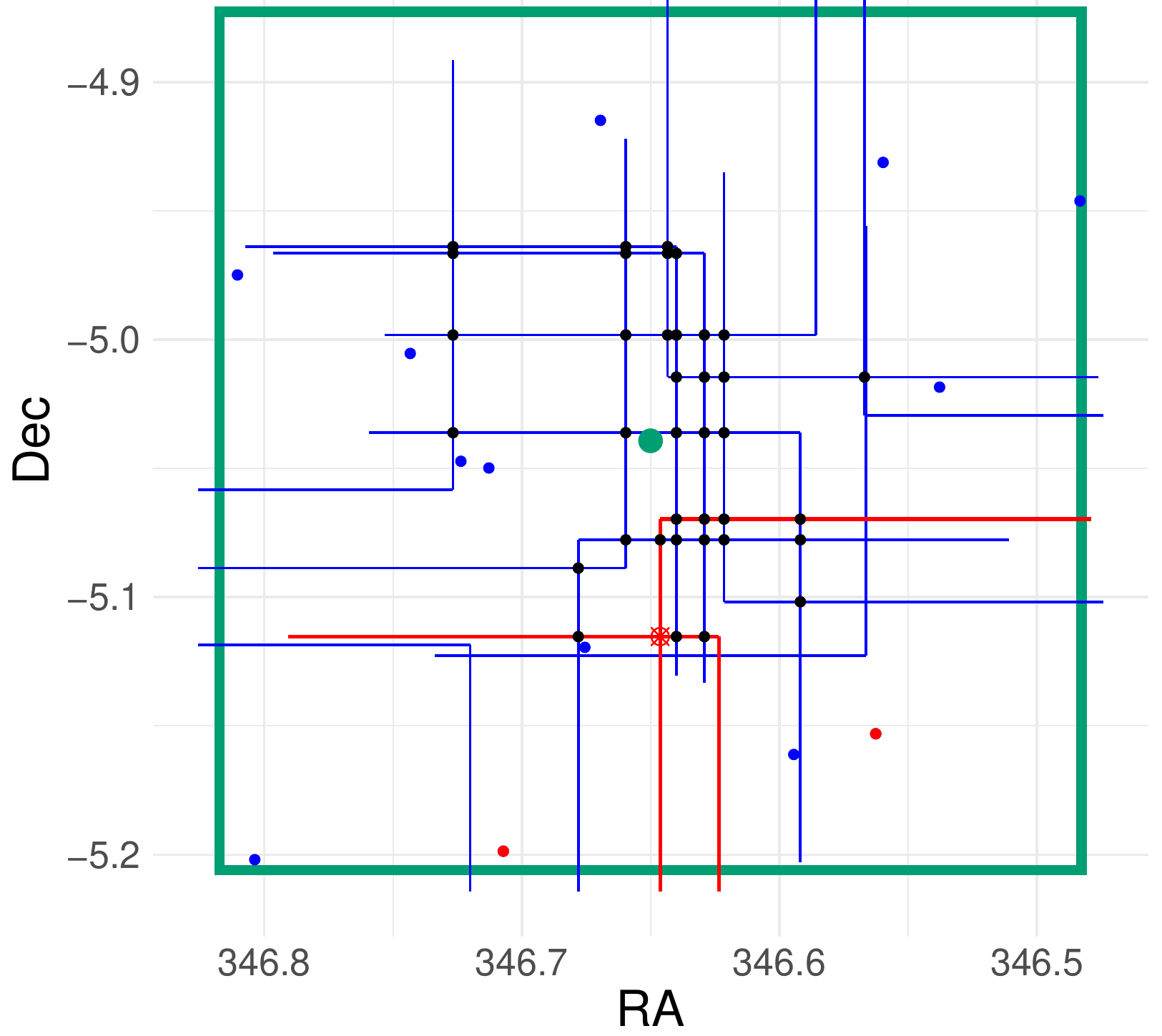}
}
\caption{\label{candidate_plot}A plot of the Candidate Zone (green box), the target (green dot), the candidate reference stars (blue dots), the effective loci for each candidate reference star (blue lines) the Points of Intersection between those loci (black dots). Highlighted in red are the candidate reference stars, loci and PoI used to identify the final pointing. Plot generated using tidyverse by \citet{tidyverse}}
\end{figure}

The locus associated with each of these 14 potential references is identified and the Points of Intersection associated with these loci are calculated. This leads to the situation shown in Figure \ref{candidate_plot}.

\begin{figure}[!htb]
\center{\includegraphics[trim={0 7.5cm 0 11cm},clip,width=0.47\textwidth]{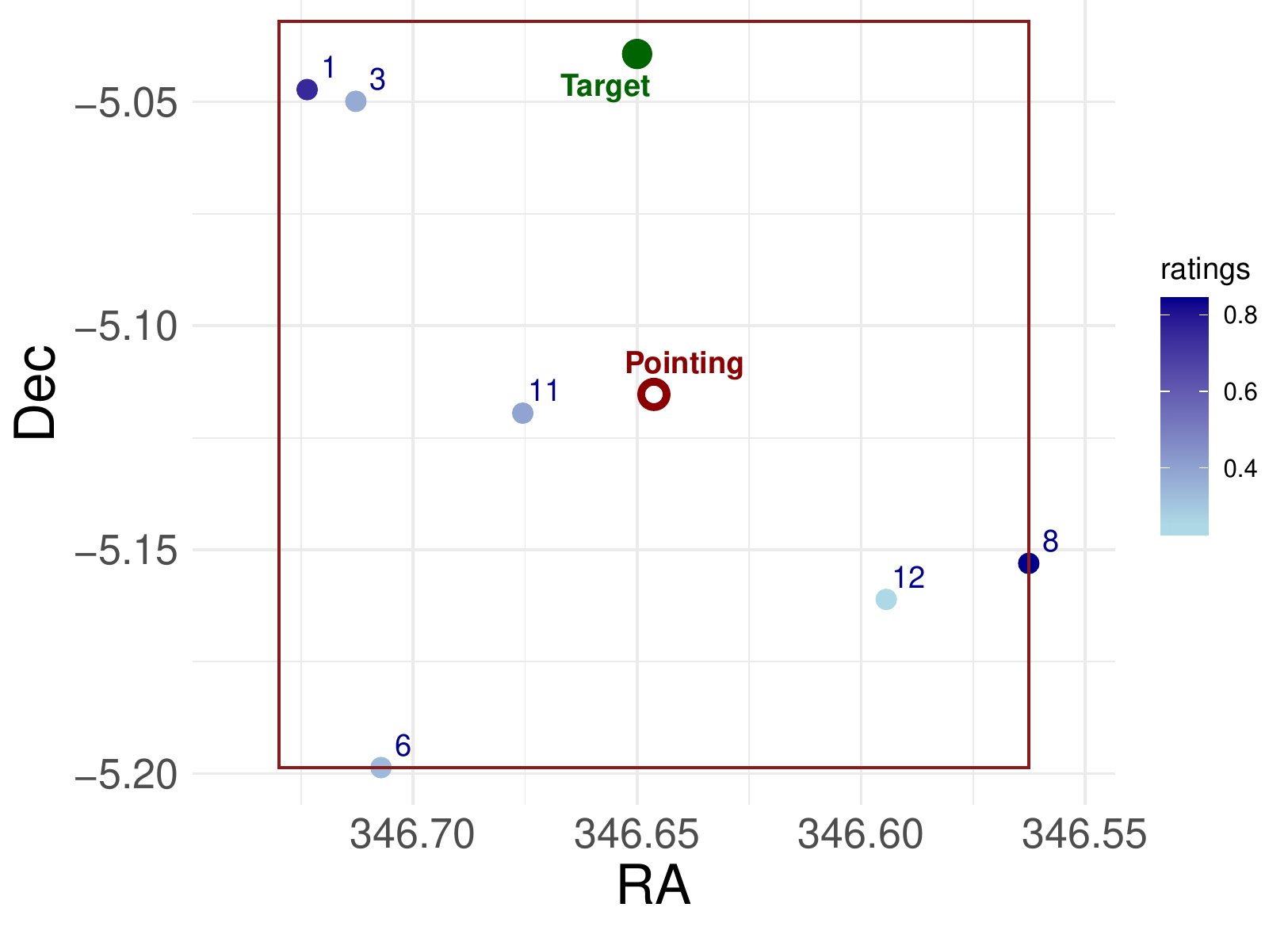}
}
\caption{\label{locus_plot}A plot illustrating the final pointing.  On this plot, the target is shown in green, the pointing and the resulting final Field of View are shown in red, and the final reference stars are shown in shades of blue that vary from light to dark based on the rating of the reference star. Plot generated using tidyverse by \citet{tidyverse}}
\end{figure}

All these 38 points of intersection are checked in turn as potential pointings. For each one, a field of view is constructed and all the potential reference stars within each one are identified. This is used to calculate a score for each point of intersection. For the star, \textit{SDSS1237680117417115655}, an optimised pointing was thus discovered with ${RA = 346.6463^{\circ} },\ {Dec = -5.1153^{\circ} }$ The field of view centred on this pointing included both the target and 7 reference stars (shown in red in Table \ref{candidate_references}). Following Expression \ref{scoring_def}, this pointing is calculated to have a score of 3.87 and is illustrated in Figure \ref{locus_plot}.

\section{Applications of the Algorithm}
\label{Applications}
The pointings generated in this algorithm are the optimum pointing for each target given the observational parameters and scoring system used. This is of use to any observer aiming to carry out differential photometry observations as it automates the selection of pointing and identification of reference stars. Two main use-cases are envisaged for this system: targeted use (where an observer wishes to identify the optimum pointing for a pre-determined target or set of targets) and catalogue generation (where many targets are submitted to the system and a set of scores and pointings are generated for each). 

\subsection{Targeted Use}
This scenario considers an observer who wishes to perform differential photometry observations of a pre-determined target(s).  The observer may use the algorithm to identify the optimum pointing for their target(s).  As illustrated for the target above, this pointing may be offset from the target but will always include the target and the reference stars with the maximum combined rating.  Software to identify this optimum pointing and select the reference stars centred on that pointing is available online at \citet{githubrepo}.  A web interface to this software is planned. 

\subsection{Catalogue Generation}
By submitting many targets at once, the optimum pointing for each can be determined and scores calculated for each.  These are output together and can be collated into a catalogue as demonstrated in \citet{quasarpaper}. The catalogues can then be used by an observer to select targets suitable for their needs by filtering the catalogue. For example, by using the scores for each target, targets with a higher score (and thus a better set of reference stars) can be selected for observation over targets with worse scores.  The catalogue production software generates lists of targets, their pointings and the scores associated with those pointings.  Users who select targets from the catalogues can then follow up their choice by identifying the set of reference stars with the SQL queries found at \citet{githubrepo}.

\section{Conclusions}
\label{Conclusions}

This paper presents the Locus Algorithm, a novel system for the identification of optimal pointings for differential photometry given a set of parameters of the planned observation provided by the user. The algorithm is presented in two stages. In the first stage, the concept of the algorithm is laid out and the steps of the algorithm are defined. In the second stage, a fully-worked example is shown applying the algorithm to the star \textit{SDSS1237680117417115655} for a 10$^\prime$ FoV. The premise of the algorithm is that a locus can be defined about any target or reference star, upon which a Field of View (FoV) can be centred and include that object at the edge of the FoV. At the Points of Intersection (PoI) between these loci, the set of targets which can be included in a FoV changes if the FoV moves in any direction. Therefore, these PoI are the essential points to compare when
determining the maximum number and quality of reference stars which can be included in a FoV of a given size.

\subsection{Summary of Algorithm}
The algorithmic description can be summarised in the following nine steps:

\textbf{Identify the target}: The target is identified by its coordinates in the RA/Dec coordinate system.

\textbf{Provide observational parameters}: The algorithm requires a FoV size, magnitude and colour difference limits and a resolution parameter to be provided.

\textbf{Define a Candidate Zone}: identify all stars which could be included in a FoV with the target by translating the position of the FoV in accordance with Expression \ref{CZdef}, which defines the Candidate Zone (CZ).

\textbf{Filter Candidate References}: for each star in the CZ, apply the filtering criteria magnitude, colour and resolution (Expressions \ref{maglim} and \ref{collim}) to identify the candidate reference stars.

\textbf{Calculate Rating}: for each candidate reference star, a rating is calculated to indicate how closely its colour matches that of the target in accordance with \ref{rating_def}.

\textbf{Calculate Loci}: The effective locus around each candidate reference star (i.e. the path upon which the FoV may be centred and include the candidate and the target) is calculated as per Expressions \ref{CPDef} and \ref{DirDef}.

\textbf{Identify Points of Intersection}: The points where the effective loci for two candidate reference stars intersect with one another are identified by combining their coordinates as shown in Expression \ref{PoICheck}.

\textbf{Calculate Score}: For each PoI, a score is calculated by combining the ratings for each candidate reference star which can be included in a FoV centred on that target in accordance with \ref{scoring_def}.

\textbf{Output Optimum Pointing}: The PoI with the best score is then selected as the optimised pointing for that target.

\subsection{Summary of Demonstration}
For demonstration purposes, this algorithm has been applied to \textit{SDSS1237680117417115655} (referred to as the target). The same nine steps are applied as outlined below

\textbf{Identify the target}: The target is found at (${RA = 346.6500^{\circ} }$, ${Dec =-5.0393^{\circ} }$).

\textbf{Provide observational parameters}: The FoV for the demonstration is 10 $'$ (0.1667$^{\circ}$ ). The magnitude difference limit is $\pm$ 2.0 mag. The colour difference limit is $\pm$0.1 mag. The resolution selected is 11$"$ (0.003$^{\circ}$ ).

\textbf{Define a Candidate Zone}: The CZ around the target is defined by (${346.4827^{\circ}  \leq RA \leq 346.8173^{\circ} {}}$, ${-5.2060^{\circ}  \leq Dec \leq -4.8726^{\circ} }$) and contains 1345 objects.

\textbf{Filter Candidate References}: A candidate reference must meet the following filtering criteria 
(${12.648 \leq r \leq 16.648}$), 
(${0.634 \leq g-r \leq 0.834}$) and 
(${0.149 \leq r-i \leq 0.349}$) and have no star 
within 0.003$^{\circ}$  of it. Applying the filtering criteria leaves 14 Candidate References.

\textbf{Calculate Rating}: for each of the candidate reference stars a rating is calculated based on the difference in colour between the target and the reference star, for example, the rating of Star \#{}8 is 0.83.

\textbf{Calculate Loci}: For each of the 14 candidate reference stars, a locus is calculated, for example for Star \#{}8, the locus is defined by ${RA_c=346.6463^{\circ} {}}$ ${Dec_c=5.0697^{\circ} {}}$, ${DirRA = -ive}$ and ${DirDec = -ive}$.

\textbf{Identify Points of Intersection}: Given 14 candidate reference stars and their corresponding Loci, there are 364 possible combinations that could be a PoI, which are generated by combining the coordinates of the corner-points of the Loci in pairs and checking the directions of the lines from each to determine whether a PoI actually exists.  PoI actually exist in 38 cases. For example, a PoI exists between the loci for Stars \#{}6 and \#{}8 at (${RA=346.6463^{\circ} }, {Dec = -5.1153^{\circ} }$).

\textbf{Calculate Score}: Score is calculated by identifying the candidate reference stars which can be included in an FoV centred on each PoI and summing their ratings. For example, for the PoI between stars \#{}6 and \#{}8 above, seven candidate reference stars can be included in the FoV and their ratings sum to 3.87.

\textbf{Output Optimum Pointing}: By comparing the scores for each PoI, the one with the highest score can be determined to be the pointing used in this example at (${RA=346.6463^{\circ} }, {Dec = -5.1153^{\circ} }$) and ${Score = 3.87}$.

\section{Further developments}
A software system has been developed to implement this algorithm and is available from \citet{githubrepo}. A paper describing the software has been submitted at \citet{locus_software_paper}. Scaling that software to allow for large-scale use required the use a Grid Computing solution, described at \citet{grid_system_paper}. By supplying a set of quasars from the 4\textsuperscript{th} Quasar Catalogue of the Sloan Digital Sky Survey (SDSS) as targets and the remainder of SDSS as potential reference stars, it was possible to generate a catalogue of optimised pointings for 26779 quasars as per \citet{quasarpaper,ZenodoQuasarCatalogue}. Using all of the stars in SDSS as targets allowed for the optimum pointing to be determined for all such stars. These pointings are of use, for example, in the search for extrasolar planets by the transit method, where high-precision differential photometry is required \citep{ZenodoXOPCatalogue}.  A paper describing the latter catalogue is in preparation.

\section*{Acknowledgements}
\textbf{Funding for this work}: This publication has received funding from Higher Education Authority Technological Sector Research Fund and the Institute of Technology, Tallaght, Dublin Continuation Fund (now Tallaght Campus, Technological University Dublin).

\textbf{SDSS Acknowledgement}: This paper makes use of data from the Sloan Digital Sky Survey (SDSS).  Funding for the SDSS and SDSS-II has been provided by the Alfred P. Sloan Foundation, the Participating Institutions, the National Science Foundation, the U.S. Department of Energy, the National Aeronautics and Space Administration, the Japanese Monbukagakusho, the Max Planck Society, and the Higher Education Funding Council for England. The SDSS Web Site is http://www.sdss.org/.

The SDSS is managed by the Astrophysical Research Consortium for the Participating Institutions. The Participating Institutions are the American Museum of Natural History, Astrophysical Institute Potsdam, University of Basel, University of Cambridge, Case Western Reserve University, University of Chicago, Drexel University, Fermilab, the Institute for Advanced Study, the Japan Participation Group, Johns Hopkins University, the Joint Institute for Nuclear Astrophysics, the Kavli Institute for Particle Astrophysics and Cosmology, the Korean Scientist Group, the Chinese Academy of Sciences (LAMOST), Los Alamos National Laboratory, the Max-Planck-Institute for Astronomy (MPIA), the Max-Planck-Institute for Astrophysics (MPA), New Mexico State University, Ohio State University, University of Pittsburgh, University of Portsmouth, Princeton University, the United States Naval Observatory, and the University of Washington.

This paper makes use of the following software packages:

\textbf{RCurl}: General Network (HTTP/FTP/...) Client Interface for R \citep{rcurl}.

\textbf{tidyverse}: R packages for data science \citep{tidyverse}.

\textbf{knitr}: A General-Purpose Package for Dynamic Report Generation in R \citep{knitr1,knitr2,knitr3}

\textbf{kableExtra}: Construct Complex Table with `kable' and Pipe Syntax \citep{kableExtra}


\bibliographystyle{elsarticle-harv}
\bibliography{Locus_Whole}

\end{document}